\documentclass[A4paper,unsortedaddress,superscriptaddress,prb,twocolumn]{revtex4-1}

\usepackage{hyperref}
\usepackage{epsfig}
\usepackage{graphicx}
\usepackage{subfigure}
\usepackage{latexsym}
\usepackage{color}
\usepackage{fullpage}
\usepackage{dcolumn}
\usepackage{bm}
\usepackage{ulem}
\usepackage{units}
\usepackage{amsmath}

\usepackage{units}

\begin{document}

%opening
\title{Topography of the graphene/Ir(111) moiré studied by surface x-ray diffraction}

\author{Fabien Jean}
\affiliation{CNRS, Inst NEEL, F-38042 Grenoble, France}
\affiliation{Univ. Grenoble Alpes, Inst NEEL, F-38042 Grenoble, France}
\author{Tao Zhou}
\affiliation{Univ. Grenoble Alpes, Inst NEEL, F-38042 Grenoble, France}
\affiliation{CEA, INAC-SP2M, Grenoble, F-38054, France}
\author{Nils Blanc}
\affiliation{CNRS, Inst NEEL, F-38042 Grenoble, France}
\affiliation{Univ. Grenoble Alpes, Inst NEEL, F-38042 Grenoble, France}
\author{Roberto Felici}
\affiliation{European Synchrotron Radiation Facility, Bo\^{i}te Postale 220, F-38043 Grenoble Cedex 9, France}
\author{Johann Coraux}
\affiliation{CNRS, Inst NEEL, F-38042 Grenoble, France}
\affiliation{Univ. Grenoble Alpes, Inst NEEL, F-38042 Grenoble, France}
\author{Gilles Renaud}
\affiliation{Univ. Grenoble Alpes, Inst NEEL, F-38042 Grenoble, France}
\affiliation{CEA, INAC-SP2M, Grenoble, F-38054, France}

\date{\today}%

\begin{abstract}

The structure of a graphene monolayer on Ir(111) has been investigated {\it in situ} in the growth chamber
 by surface x-ray diffraction including the specular rod,
 which allows disentangling the effect of the sample roughness from that of the nanorippling of graphene and iridium
 along the moiré-like pattern between graphene and Ir(111).
 Accordingly we are able to provide precise estimates of the undulation associated with this nanorippling,
 which is small in this weakly interacting graphene/metal system and thus proved difficult to assess in the past.
 The nanoripplings of graphene and iridium are found in phase, i.e. the in-plane position of their height maxima coincide,
 but the amplitude of the height modulation is much larger for graphene (\(0.379 \pm 0.044\) \AA) than, {\it e.g.}, for the topmost Ir layer (\(0.017 \pm 0.002\) \AA).
 The average graphene-Ir distance is found to be \(3.38 \pm 0.04\) \AA.

\end{abstract}

\maketitle

Graphene, a monoatomic layer of carbon atoms arranged in a honeycomb lattice,
 has been investigated thoroughly in the past ten years because of its exceptional properties, which hold promises for numerous applications.\cite{Novoselov2012}
 Transition metal surfaces form a broad family of substrates for the growth of large area, high quality graphene.\cite{Tetlow2014}
 New properties can be induced in graphene through the interaction with the substrate,
 e.g. electronic bandgaps,\cite{Pletikosic2009} spin-polarization \cite{Varykhalov2008} and superconductivity.\cite{Tonnoir2013}
 In most graphene-on-metal systems, the interaction is modulated at the nanoscale,
 due to lattice mismatch between graphene and the metal,
 which results in two-dimensional patterns with periodicity of the order of nanometers,
 often referred to as "moirés", following an analogy with the beating of optical waves through two mismatched periodic lattices ({\it e.g.} tissue veils).
 Knowledge on the topographic properties of these moirés, {\it i.e.} the average graphene-metal distance,
 and the perpendicular-to-the-surface amplitude of the graphene and metal undulations across the moiré,
 is desirable in view of characterizing the interaction and rationalizing the other properties.

%%% Figure Reciprocal Space Map

\begin{figure}[h]
\begin{center}
 \includegraphics[width=150pt]{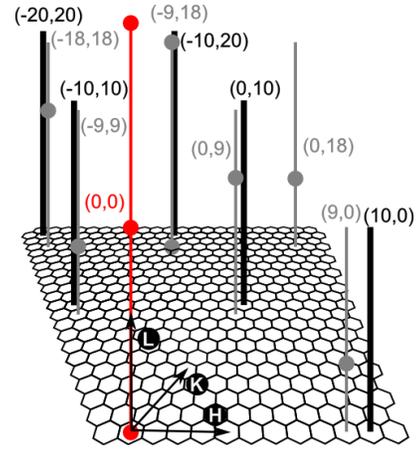}

\caption{Sketch of the reciprocal space, the hexagonal grid shows the partition of its (\textit{H},\textit{K}) plane according to the 10-on-9 commensurability.
 \textit{H}, \textit{K} and \textit{L} are in reciprocal lattice unit of the moiré (superlattice) surface unit cell.
 In gray are shown the measured CTRs from the iridium, with circles to highlight the positions of the different Bragg reflections.
 The graphene rods are shown in black.
 The specular CTR (\textit{H}=\textit{K}=0) is shown in red.
 Each is labeled with its (\textit{H},\textit{K}) position in the 10-on-9 moiré surface supercell.
 }
\label{Map RR}
\end{center}
\end{figure}

%%% Figure Data CTRs + reflectivity

\begin{figure}[h]
\begin{center}
 \includegraphics[width=200pt]{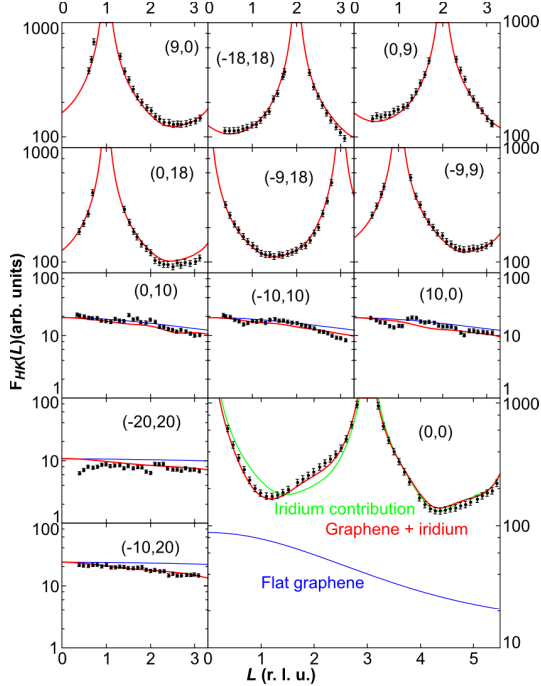}

\caption{Experimental structure factors \( F_{H,K}(L) \) of iridium CTRs and graphene rods from SXRD measurements of the first sample in black with the error bars.
 The solid red lines represent the best fit with the Fourier model.
 In blue with the rods is the contribution of a flat graphene layer alone, to highlight the effect of the roughness and undulations on the rods.
 The specular rod (0,0) from the second sample is reported in the bottom right in black with error bars.
 The solid red line represents the final fit, the green one the contribution of the iridium alone and in blue the contribution of a flat graphene layer alone.
 }
\label{Graphe CTRs}
\end{center}
\end{figure}

The topography is however hard to grasp at such small scales.
 Most efforts that have relied on scanning tunnelling microscopy have faced the issue of the entanglement of the structural
 and local density of state which is inherent to the tunnelling effect.
 A striking illustration has been the debate on the amplitude \cite{Marchini2007, VazquezdeParga2008}
 and sign \cite{Busse2011, Sun2011} of the moiré-related undulation in graphene/Ru(0001) and graphene/Ir(111), respectively.
 Atomic force microscopy has proven to be a valuable alternative, provided it is performed with care,
 especially with respect to the possible chemical interaction between tip and sample.\cite{Dedkov2014, Hamalainen2013}
 Scattering techniques, such as low-energy electron diffraction (LEED), surface X-ray diffraction (SXRD), and X-ray standing waves (XSW),
 are free of such probe-induced perturbations of the systems.
 To the expense of complex calculations in the framework of the dynamical theory of diffraction,
 LEED was used to assess the topography of graphene/Ru(0001) \cite{Moritz2010} and graphene/Ir(111).\cite{Hamalainen2013}
 SXRD was used to analyze the topography of graphene/Ru(0001),\cite{Martoccia2008} as was done by XSW for graphene/Ir(111).\cite{Busse2011}
 Confirming and refining the results obtained with these approaches is of prime importance in order to set reliable points of reference for first principle calculations,
 which are cumbersome in essence in such systems due to the importance of non-local ({\it e.g.} van der Waals) interactions.\cite{Mittendorfer2011}

Here, we address the model graphene/Ir(111) system, typical of a weak graphene-metal interaction.
 Its moiré topography only slightly deviates from the flat case and is thus difficult to characterize.
 With the help of two techniques, SXRD and extended x-ray reflectivity (EXRR),
 the latter not having been employed to characterize monolayer graphene on a substrate before,
 we deduce an average \(3.38 \pm 0.04\) \AA \space distance between graphene and Ir(111), and determine,
 with an uncertainty as low as with scanning probe microscopy,\cite{Hamalainen2013} a \(0.379 \pm 0.044\) \AA \space amplitude of the graphene undulation.
 Besides, we are able to estimate the undulation of the Ir layers, which is usually not accessible to other techniques, \(0.017 \pm 0.002\) \AA \space for the topmost one.

The synchrotron x-ray diffraction measurements were performed in ultra-high vacuum chambers
 coupled with \textit{z} axis diffractometers at the BM32 and ID03 beamlines of the European Synchrotron Radiation Facility.
 Details on the chambers and the beam are given in Ref.~\onlinecite{Jean2013}.
 The non-specular crystal truncation rods (CTRs) were measured on BM32 and the specular rod, 00{\it L}, was measured by EXRR on ID03.
 The x-ray beam energy was set at 11 keV.
 The reciprocal space scans of the scattered intensity presented below are all normalized to the intensity measured with a monitor placed before the sample.
 For the SXRD measurements, the intensity along the Ir(111) crystal truncation rods (CTRs) and along the graphene rods was measured with a Maxipix two-dimensional detector
 in stationary mode for the upper range of the out-of-plane scattering vector component ({\it i.e.} large values of the out-of-plane reciprocal space coordinate {\it L}),
 and by performing sample rocking scans for low {\it L}-values.\cite{Drnec2014}
 The amplitude of the structure factors $F_{H,K}({\it L})$ - the square root of the measured intensity - for the different CTRs and graphene rods,
 corresponding each to different values of the in-plane reciprocal space parameters {\it H} and {\it K},
 were extracted and processed with the PyRod program described in Ref. \onlinecite{Drnec2014}.
 PyRod was also used to simulate the structure factors using the model described below,
 and to refine the structural parameters of this model with the help of a least squares fit of the simulation to the data.
 The total uncertainty on the experimental structure factors is dominated by the systematic error estimated to be 6.1\%, according to Ref. \onlinecite{Drnec2014};
 the statistical error being everywhere smaller than 1 \%.

The Ir single crystals were cleaned according to a procedure described in Ref. \onlinecite{Jean2013} allowing for considerably reducing the concentration of residual carbon in bulk Ir(111).
 Graphene was grown in two steps, first by 1473 K annealing of a room-temperature-adsorbed monolayer of ethylene, second by exposure to 10$^{-8}$ mbar of ethylene the surface held at 1273 K.
 This growth procedure allows for selecting a well-defined crystallographic orientation of purely single-layer graphene with respect to Ir(111).\cite{vanGastel2009}
 Compared to the samples studied in Refs. \onlinecite{Busse2011,Hamalainen2013}, the surface coverage is larger (100\%) in our case.
 Our growth procedure is similar to that used to prepare the 100\%-coverage graphene studied in Ref. \onlinecite{Runte2014}, yet the temperatures which we chose for each step are different,
 actually identical to those used for preparing one of the samples addressed in Ref. \onlinecite{Jean2013}.
 We note that both the graphene coverage and growth temperature have been argued to influence the structure of graphene,\cite{Busse2011,Blanc2012} and thus its properties.\cite{Usachov2012}
 Two samples were prepared, one in each of the UHV chambers installed at the BM32 and ID03 beamlines where the SXRD and EXRR experiments were performed respectively.
 The hexagonal lattice unit cell of the iridium surface has a lattice parameter of 2.7147 \AA \space at room temperature.
 The graphene unit cell has a measured lattice parameter of 2.4530 \AA.
 The ratio between the two lattice parameters, 0.903, is close to 0.9.
 Therefore, in the following we assume that the system is commensurate,
 with a \( (10 × 10) \) graphene cell coinciding with a \( (9 × 9) \) iridium one.
 In the following, the in-plane unit cell of reciprocal space is the moiré one.\cite{Coraux2008}
 This corresponds to {\it H} or {\it K} indexes multiples of 9 and 10 for Ir CTRs and graphene rods, respectively (Fig. \ref{Map RR}).

Figure \ref{Graphe CTRs} shows the Ir CTRs and graphene rods.
 As expected for a (essentially) two-dimensional layer such as graphene, the graphene rods are basically featureless.\cite{Charrier2002}
 Qualitatively, because the undulations of the graphene and top substrate layers are expected to be small,
 the main features are {\it i}) the pronounced interference effect on the specular rod $F_{0,0}({\it L})$
 related to the average distance $dz_{Gr}$ between Ir and graphene, expected to be larger than the bulk distance of 2.2 \AA;
 {\it ii}) the decrease of the otherwise featureless CTRs in between Bragg peaks, related to the substrate roughness ;
 and {\it iii}) the decrease of the graphene rods with increasing {\it L}, dominated by the undulation of the graphene layer,
 as shown with the simulated graphene rods for a flat graphene layer alone in Fig.  \ref{Graphe CTRs}.
 This decorrelation between roughness and undulation allows these parameters to be determined with high accuracy.

%%% Figure Parameters

\begin{figure}[h]
\begin{center}
 \includegraphics[width=160pt]{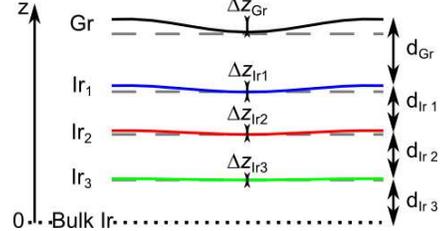}

\caption{Sketch of the parameters studied. In black is the graphene and in blue, red and green are the surface layers of iridium.
 The amplitudes of their corrugation are shown by arrows in the middle.
 The start of the bulk iridium is sketched with the dotted black line.
 The gray dashed lines represents the expected bulk positions for the different atomic layers without corrugation.
 The z-axis on the left is a reference to the linear dependency of the iridium corrugation amplitude.}
\label{Param}
\end{center}
\end{figure}

\begin{table*}[h]
\begin{center}
\begin{tabular}{|r@{}r@{}r@{}r@{}r@{}r|}
\hline
\(A^x_{s,t}=\)&\(-A^x_{t,-s-t} + A^y_{t,-s-t} = \) & \( -A^y_{-s-t,s} = \) & \( -A^y_{-t,-s} = \) & \( -A^x_{-s,s+t} + A^y_{-s,s+t} = \) & \( A^x_{s+t,-t}\)\\ \hline
\(A^y_{s,t}=\)&\(A^x_{-s-t,s} - A^y_{-s-t,s} = \) & \( -A^x_{t,-s-t} = \) & \( -A^x_{-t,-s} = \) & \( -A^y_{-s,s+t} = \) & \( A^x_{s+t,-t} -A^y_{s+t,-t}\)\\ \hline
\(A^z_{s,t}=\)&\(A^z_{t,-s-t} = \) & \( A^z_{-s-t,s} = \) & \( A^z_{-t,-s} = \) & \( A^z_{-s,s+t} = \) & \( A^z_{s+t,-t}\)\\ \hline \hline
\(B^x_{s,t}=\)&\(-B^x_{t,-s-t} + B^y_{t,-s-t} = \) & \( -B^y_{-s-t,s} = \) & \( -B^y_{-t,-s} = \) & \( -B^x_{-s,s+t} + B^y_{-s,s+t} = \) & \( B^x_{s+t,-t}\)\\ \hline
\(B^y_{s,t}=\)&\(B^x_{-s-t,s} - B^y_{-s-t,s} = \) & \( -B^x_{t,-s-t} = \) & \( -B^x_{-t,-s} = \) & \( -B^y_{-s,s+t} = \) & \( B^x_{s+t,-t} - B^y_{s+t,-t}\)\\ \hline
\(B^z_{s,t}=\)&\(B^z_{t,-s-t} = \) & \( B^z_{-s-t,s} = \) & \( B^z_{-t,-s} = \) & \( B^z_{-s,s+t} = \) & \( B^z_{s+t,-t}\)\\ \hline
\end{tabular}
\end{center}
\caption{Relationships between the Fourier coefficients \(A^i_{s,t}\) and \(B^i_{s,t}\) \((\it{i} = \{x, y, z\})\) }
\label{table Fourier}
\end{table*}

In order to achieve a quantitative characterization of the topography of the system, we introduce a simple model.\cite{Note1}
 A limited set of parameters (see Fig. \ref{Param}), including the average interplanar distances, the actual roughness, and the amplitude of undulation of each layer,
 seems to be a reasonable option for a simple modeling of the system.
 In order to approach this description, we introduce a lattice model based on a Fourier series, such as the one proposed for graphene/Ru(0001).\cite{Martoccia2010}
 In this model, the displacement in the direction \textit{i} ({\(\textit{i} =\)} \{{\it x,y,z}\}) of an atom with {\it x}, {\it y} and {\it z} coordinates,
 with respect to the corresponding position in a flat layer, is given by

\begin{equation}
\mathrm d r^i = \sum_{s,t} A^i_{s,t} \times \mathrm{sin}[2 \pi (sx + ty) ] + B^i_{s,t} \times \mathrm{cos}[2 \pi (sx + ty) ]
\end{equation}

where the sum runs over the different orders of the series.
 Due to the crystal symmetry of graphene and Ir(111), the displacements must respect a \textit{p}3\textit{m}1 symmetry,
 {\it i.e.} they must fulfill

\begin{equation}
\mathrm{R_j}^{-1}\{\mathrm{d}\textbf{r}[\mathrm{R_j}(\textbf{r})]\}=\mathrm{R_j}\{\mathrm{d}\textbf{r}[\mathrm{R_j}^{-1}(\textbf{r})]\}
\end{equation}

with \textit{j} $\in$ [0,5].
 R\({_0}\) is the identity matrix, R\({_1}\) and R\({_2}\) correspond to the $\pm$ 120° rotations and the last three to the mirror planes.

\begin{widetext}

\begin{equation}
\mathrm{R}_0 =  
\left( \begin{array}{cc}
1 & 0\\
0 & 1 \end{array} \right),
\mathrm{R}_1 =  
\left( \begin{array}{cc}
0 & \bar{1}\\
1 & \bar{1} \end{array} \right),
\mathrm{R}_2 =  
\left( \begin{array}{cc}
\bar{1} & 1\\
\bar{1} & 0 \end{array} \right),\\
\mathrm{R}_3 =  
\left( \begin{array}{cc}
0 & \bar{1}\\
\bar{1} & 0 \end{array} \right),
\mathrm{R}_4 =  
\left( \begin{array}{cc}
\bar{1} & 1\\
0 & 1 \end{array} \right),
\mathrm{R}_5 =  
\left( \begin{array}{cc}
1 & 0\\
1 & \bar{1} \end{array} \right)
\end{equation}

\end{widetext}

These symmetry constraints impose that not all Fourier coefficients in Eq. (1) are independent.
 Their relationships are given in Table \ref{table Fourier}.

In the following we further simplify the model by limiting the Fourier development to first order,
 which is legitimate due to the fact that no significant diffraction data is measurable beyond first order
 (a diffraction experiment is actually a measurement of the Fourier transform of the electronic density,
 thus, to a good approximation, of the shape of graphene).
 Besides, we assume that the undulations of all Ir layers are in phase, as found in density functionnal theory (DFT) calculations.
 In this framework, the {\it x}, {\it y} and {\it z} displacements simply write:

\begin{widetext}

\begin{equation}
\begin{aligned}
\mathrm{d}r^x = {} & A^x \times (2 \times \mathrm{sin}(2 \pi x) + \mathrm{sin}(2 \pi y) + \mathrm{sin}(2 \pi (x-y)))\\
 &+ B^x \times (2 \times \mathrm{cos}(2 \pi x) - \mathrm{cos}(2 \pi y) - \mathrm{cos}(2 \pi (x-y)))
\end{aligned}
\end{equation}

\begin{equation}
\begin{aligned}
\mathrm{d}r^y = {} & A^x \times (\mathrm{sin}(2 \pi x) + 2 \times \mathrm{sin}(2 \pi y) - \mathrm{sin}(2 \pi (x-y)))\\
&+ B^x \times (\mathrm{cos}(2 \pi x) - 2 \times \mathrm{cos}(2 \pi y) + \mathrm{cos}(2 \pi (x-y)))
\end{aligned}
\end{equation}

\begin{equation}
\begin{aligned}
\mathrm{d}r^z = {} & A^z \times (2 \times \mathrm{sin}(2 \pi x) - 2 \times \mathrm{sin}(2 \pi y) - 2 \times \mathrm{sin}(2 \pi (x-y)))\\
& + B^z \times (2 \times \mathrm{cos}(2 \pi x) + 2 \times \mathrm{cos}(2 \pi y) + 2 \times \mathrm{cos}(2 \pi (x-y)))
\end{aligned}
\end{equation}

\end{widetext}

 Thus, only two variables per atomic plane, \(A^x\) and \(B^x\), are needed to describe the in-plane displacements.
 The model is applied to graphene/Ir(111), with three iridium layers and one graphene layer.
 Each of these layers is characterized by four Fourier coefficients (\(A^x\), \(B^x\), \(A^z\) and \(B^z\)),
 plus another parameter corresponding to an average \textit{z} displacement of the layer from its equilibrium position in the bulk.
 This distance between metal planes parallel to the surface is known to vary,
 in some cases by as much as few percents and in a non monotonous manner across the few topmost layers of metal surfaces \cite{Adams1985}.
 In order to reduce the number of free parameters however, we assume a linear dependence of the distance between Ir(111) planes, thus of of \(A^z\) and \(B^z\)), as a function of depth.
 This assumption complies with the results of the DFT simulations (cf. Table \ref{table parameters}, where the topographic parameters of the model are listed).

\begin{table*}[h]

\begin{center}
\begin{tabular}{|p{0.7cm}|p{2.9cm}|p{2.9cm}|p{2.7cm}|p{2.7cm}|p{2.7cm}|}
\hline
 & SXRD ($1^{st}$ sample) & EXRR ($2^{nd}$ sample) & DFT & Ref. \onlinecite{Busse2011} & Ref. \onlinecite{Hamalainen2013}\\\hline
\(dz_{Gr}\) & $3.39 \pm 0.28$ & $3.38 \pm 0.04$ & $3.43$ & $3.38 \pm 0.04$ & $3.39 \pm 0.03$\\\hline
\(\Delta z_{Gr} \) & $0.379 \pm 0.044$ &  & $0.46$ & $0.6 \pm 0.1$ & $0.47 \pm 0.05$\\
 &  &  &  & $1.0 \pm 0.2$ & \\\hline
\(dz_{Ir_{1}}\) & $2.203 \pm 0.012$ & $2.203 \pm 0.010$ & $2.190$ &  & $2.222$\\
\(dz_{Ir_{2}}\) & $2.212 \pm 0.007$ & $2.205 \pm 0.008$ & $2.175$ &  & $2.224$\\
\(dz_{Ir_{3}}\) & $2.223 \pm 0.002$ & $2.225 \pm 0.004$ & $2.184$ &  & $2.222$\\\hline
\(\Delta z_{Ir_{1}} \) & \(0.017 \pm 0.002\) &  & $0.015$ &  & $0.006$\\
\(\Delta z_{Ir_{2}} \) & \(0.011 \pm 0.001\) &  & $0.012$ &  & $0.006$\\
\(\Delta z_{Ir_{3}} \) & \(0.006 \pm 0.001\) &  & $0.004$ &  & $0$\\\hline
$\rho$ & \(0.42 \pm 0.20\) & \(1.05 \pm 0.08\) &  &  & \\\hline
\(O_{Gr}\) & $98 \pm 2$\% & $89.7 \pm 1$\% & $100$\% & $39$\% & Partial\\
 &  &  &  & $63$\% & \\\hline
\end{tabular}
\end{center}
\caption{Topographic parameters for the two samples, the DFT calculations data from Ref. \onlinecite{Busse2011}
 and results from Ref. \onlinecite{Busse2011} (XSW) and \cite{Hamalainen2013} (LEED + AFM).
 \(dz_{Gr}\) is the mean distance between the graphene and its substrate;
 \(\Delta z_{Gr}\) is the graphene undulation amplitude;
 \(dz_{Ir_{1}}\), \(dz_{Ir_{2}}\) and \(dz_{Ir_{3}}\) are the interlayer distances of the iridium surface layers and
 \(\Delta z_{Ir_{1}} \), \(\Delta z_{Ir_{1}} \) and \(\Delta z_{Ir_{1}} \) are their undulation amplitudes;
 $\rho$ is the roughness of the sample surface;
 \(O_{Gr}\) is the graphene coverage in percent.
 All the parameters are in ångströms (\AA) except the coverage.}
\label{table parameters}

\end{table*}

\begin{figure}[h]
\begin{center}
 \includegraphics[width=200pt]{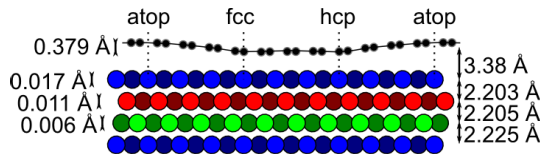}

\caption{Cut of the 10-on-9 commensurability to represent the corrugations and displacements of the atomic layers.
 Carbon atoms of the graphene are black circles, the iridium atoms are in blue, red and green to show the ABC stacking of the different layers.
 The three coincidence regions of graphene with the substrate as well as the corrugations and interlayer spacing are denoted.}
\label{Final}
\end{center}
\end{figure}

The Fourier model was used to fit the SXRD data.
 The expected in-plane displacements (Fig.~\ref{Final}), below 0.01 \AA \space according to first principle calculations,\cite{Busse2011}
 have no noticeable influence on the Ir CTRs and graphene rods, and are discarded in the simulations.\cite{Note2}
 The best fit lead to a \(\chi^{2}\) value of 3.5 and the results are shown in Table \ref{table parameters}.
 We find a 98 $\pm$ 2\% graphene coverage.
 The graphene is found to have a mean distance of \(dz_{Gr} = 3.39 \pm 0.28\) \AA \space with its substrate and a corrugation of \(\Delta z_{Gr} = 0.379 \pm 0.044\) \AA.
 The graphene distance with its substrate is close to the interlayer spacing in graphite, 3.36 \AA.
 As explained above, the benefit of the SXRD analysis of both graphene and Ir contributions is to provide an accurate value of the amplitude
 of the graphene undulation perpendicular to the surface, as compared to other techniques.
 The interlayer Ir spacings are found to be \(2.203 \pm 0.012\) \AA, \(2.212 \pm 0.007\) \AA \space and \(2.223 \pm 0.002\) \AA \space from top to bottom.
 The topmost layer of iridium has an undulation of \(0.017 \pm 0.002\) \AA, the second layer has an undulation of \(0.011 \pm 0.001\) \AA,
 and the last one is \(0.006 \pm 0.001\) \AA.
 Finally, the roughness of the iridium substrate is found to be \(0.42 \pm 0.20\) \AA, following a simple $\beta$-model. \cite{Robinson1986}
 This small value may be linked with the small coherence length of the X-ray beam (corresponding to about 10 flat Ir terraces separated by atomic step edges) on the BM32 beamline.

%%% Figure 10 on 9 + Final positions

The best fit between simulations and SXRD data is achieved for an iridium undulation in phase with the graphene one,
 with a smaller amplitude though.
 This finding is at variance with that obtained in earlier scanning probe microscopy measurements performed in specific imaging conditions,\cite{Sun2011}
 and supports the picture progressively assembled through other reports, based on scanning probe microscopies,\cite{Dedkov2014, Boneschanscher2012}
 XSW,\cite{Busse2011} and first principle calculations.\cite{Ndiaye2006, Busse2011}

%%% Figure Fourier model on DFT

\begin{figure*}[h]
\begin{center}
 \includegraphics[width=350pt]{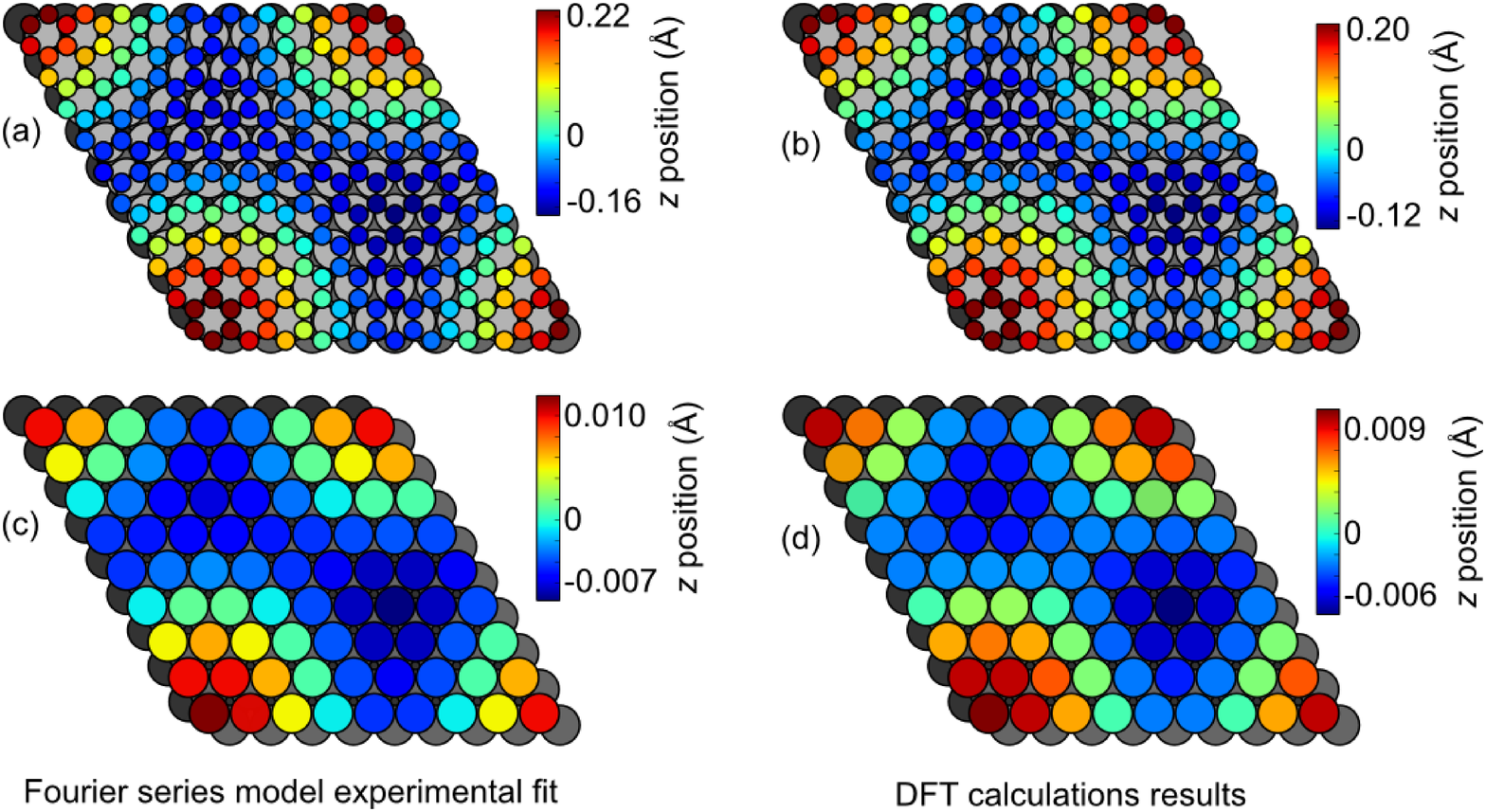}

\caption{Sketches of the graphene out of plane variations from (a) the Fourier model used on the experimental data and (b) the DFT calculations results
 and the out of plane variations of the topmost iridium layer from (c) the Fourier model used on the experimental data and (d) the DFT calculations results.
 with the DFT calculations results shown in left half-discs and the Fourier series fit in right half-discs
 The out of plane corrugation is shown with a color gradient, with the scales in \AA.}
\label{Fourier DFT}
\end{center}
\end{figure*}

The main limitation of this SXRD analysis is the rather large uncertainty on the $dz_{Gr}$ distance.
 This motivated complementary measurements of the specular rod on the second sample, using the ID03 setup
 as the extended reflectivity was not accessible in the BM32 setup.
 The EXRR result is shown in Fig. \ref{Graphe CTRs} together with the best fit and simulated and graphene specular rods.
 The best fit of the specular rod, yielding a \(\chi^{2}\) value of 1.064, was done with a simplified model,
 in which the undulations of both the iridium or graphene were fixed at the values obtained from the SXRD analysis.
 It yields a value of  $dz_{Gr}$ = 3.38 \AA, very close to that determined on the other sampler by off-specular SXRD, but with a much better accuracy, $\pm$ 0.04 \AA.
 The graphene layer of this second sample is found incomplete, with a 90 $\pm$ 2\% graphene coverage.
 In addition, the spacings between the topmost Ir planes, found to be 2.203 $\pm$ 0.010 \AA,
 2.205 $\pm$ 0.008 \AA \space and 2.225 $\pm$ 0.004 \AA \space from top to bottom (2.217 \AA \space in the bulk).
 We find interlayer distances between Ir(111) planes that are larger than those found in the absence of graphene.\cite{He2008}
 This finding is consistent with the $p$-doping found for graphene\cite{Pletikosic2009} which implies electron transfers from graphene to Ir(111).
 The corresponding higher electronic density in Ir(111) is expected to counterbalance the surface relaxation in bare Ir(111).
 The substrate roughness in this case is found to be 1.1 $\pm$ 0.1 \AA, larger than that obtained from SXRD.
 This is however expected since the coherence length of the beam is two orders of magnitude larger here,
 \textit{e.g.} around 1000 atomic steps of the substrate scatter the beam coherently.

This is the first study of a sample with a complete graphene coverage,
 thus the deviations from previous studies can be explained due to strains in the full layer that can relaxe in graphene island.
 This could also be explained by the difference in the growth process (temperature, methods...)
 and Busse \textit{et al.} \cite{Busse2011} showed that the undulation varies depending on the graphene coverage.
 Moreover, the undulation could also be affected by the growth methods (full/partial growth, chemical vapor deposition, temperature programmed growth...)
 and growth temperature as it has been reported that these parameters affect the graphene lattice parameter and its commensurability with the substrate.\cite{Blanc2012, Jean2013}
 The iridium undulations are also found larger than those deduced from a LEED study.\cite{Hamalainen2013}
 This might be due to some limitation of LEED to analyse layers below the graphene one, because of the small electron mean free path.

The graphene-metal distance which we obtain is in good agreement with values deduced by XSW, LEED, and AFM (see Table \ref{table parameters}).
 The undulation of the graphene which we obtain is also in agreement with that found by LEED and AFM.
 It is however smaller than that deduced from XSW.
 The difference might originate from two effects.
 First, we recently found that the in-plane lattice parameter of graphene varies as a function of the preparation method,
 which is different in Refs. \onlinecite{Busse2011, Hamalainen2013}, several TPG cycles at 1420 K for different coverages and one TPG at 1500 K respectively,
 and in the present work, TPG at 1473 K followed by CVD at 1273 K for complete coverage.
 Given that the strain is closely related to the graphene buckling (undulation),\cite{Runte2014} we indeed expect different undulations in each of these reports.
 Second, the strain (and thus buckling) of graphene was argued to depend on the fraction of edge atoms in graphene, i.e. on graphene coverage.\cite{Busse2011}
 Our results, unlike those in Refs. \onlinecite{Busse2011, Hamalainen2013}, address close-to-full layer graphene.

The Fourier model was also tested to fit the displacements obtained by the DFT calculations described in Ref. \onlinecite{Busse2011}.
 The model was in very good accordance with the DFT calculations results, in particular the iridium top layer and graphene,
 thus confirming that the first order Fourier component is enough to describe the system, as shown in Fig \ref{Fourier DFT}.
 Moreover, it also confirmed that \(A^z\) and \(B^z\) of the three iridium layers have an almost linear dependence as a function of depth.
 From the DFT simulation, the corrugations of the iridium surface layers from top to bottom are 0.015 \AA, 0.012 \AA \space and 0.04 \AA \space while the graphene one is 0.35 \AA,
 which are close to the experimental results.

In fact, this analysis has a limite too, as our starting hypothesis on the structure of the supercell,
 a \( (10 × 10) \) graphene cell coinciding with a \( (9 × 9) \) iridium, may have an impact on the results.
 It was reported previously that this system cannot be consider fully commensurate, as it is really a composition of commensurate domains with incommensurate boundaries \cite{Blanc2012}
 and that the thermal history of the sample effects it.\cite{Blanc2012, Jean2013}
 Here, the 9.03 ratio indicates that there should be a combinaison of \( (10 × 10) \)/\( (9 × 9) \), \( (21 × 21) \)/\( (19 × 19) \) and incommensurate domains.
 However, despite the complexity of the sample, the starting hypothesis of the problem allows to extract a good approximation of the actual structure.

To conclude we have employed SXRD to determine with high resolution, on the basis of a simple structural model,
 the structure of a weakly scattering atomically thin membrane, graphene, in weak interaction with a metallic substrate made of strong scatterers, Ir atoms.
 We determine the undulation of graphene across the moiré-like superstructure formed between graphene and Ir(111), \(0.379 \pm 0.044\) Å without the ambiguity inherent to other ensemble-averaging techniques.
 Our determination of the average graphene-Ir(111) distance is consistent with previous reports based on local-probe and ensemble-averaging analysis.
 Finally we unveil the faint corrugations predicted by DFT calculation in the substrate, which are as low as \(0.017 \pm 0.002\) Å for the topmost Ir layer,
 and are characteristic of a weak C-Ir bonding having a slight covalent character in some of the sites of the moiré.
 The use of SXRD for other two-dimensional membranes, such as transition metal dichalcogenides, boron nitride, or monolayer silica,
 should allow for constructing a comprehensive picture of the nanomechanics of atomically thin membranes under the influence of substrates.

%%%%%%% Acknowledgements %%%%%%%
% \begin{acknowledgments}
We thank Olivier Geaymond, Thomas Dufrane and the staff members of the ID03 and BM32 beamlines,
 Nicolae Atodiresei for the DFT calculations data and the French Agence Nationale de la Recherche for funding (Contract No. ANR-2010-BLAN-1019-NMGEM).
% \end{acknowledgments}

%%%%%%% Bibliography %%%%%%%%%



\begin{thebibliography}{0}%
\makeatletter
\providecommand \@ifxundefined [1]{%
 \@ifx{#1\undefined}
}%
\providecommand \@ifnum [1]{%
 \ifnum #1\expandafter \@firstoftwo
 \else \expandafter \@secondoftwo
 \fi
}%
\providecommand \@ifx [1]{%
 \ifx #1\expandafter \@firstoftwo
 \else \expandafter \@secondoftwo
 \fi
}%
\providecommand \natexlab [1]{#1}%
\providecommand \enquote  [1]{``#1''}%
\providecommand \bibnamefont  [1]{#1}%
\providecommand \bibfnamefont [1]{#1}%
\providecommand \citenamefont [1]{#1}%
\providecommand \href@noop [0]{\@secondoftwo}%
\providecommand \href [0]{\begingroup \@sanitize@url \@href}%
\providecommand \@href[1]{\@@startlink{#1}\@@href}%
\providecommand \@@href[1]{\endgroup#1\@@endlink}%
\providecommand \@sanitize@url [0]{\catcode `\\12\catcode `\$12\catcode
  `\&12\catcode `\#12\catcode `\^12\catcode `\_12\catcode `\%12\relax}%
\providecommand \@@startlink[1]{}%
\providecommand \@@endlink[0]{}%
\providecommand \url  [0]{\begingroup\@sanitize@url \@url }%
\providecommand \@url [1]{\endgroup\@href {#1}{\urlprefix }}%
\providecommand \urlprefix  [0]{URL }%
\providecommand \Eprint [0]{\href }%
\providecommand \doibase [0]{http://dx.doi.org/}%
\providecommand \selectlanguage [0]{\@gobble}%
\providecommand \bibinfo  [0]{\@secondoftwo}%
\providecommand \bibfield  [0]{\@secondoftwo}%
\providecommand \translation [1]{[#1]}%
\providecommand \BibitemOpen [0]{}%
\providecommand \bibitemStop [0]{}%
\providecommand \bibitemNoStop [0]{.\EOS\space}%
\providecommand \EOS [0]{\spacefactor3000\relax}%
\providecommand \BibitemShut  [1]{\csname bibitem#1\endcsname}%
\let\auto@bib@innerbib\@empty
%</preamble>
\end{thebibliography}%


\begin{thebibliography}{30}%
\makeatletter
\providecommand \@ifxundefined [1]{%
 \@ifx{#1\undefined}
}%
\providecommand \@ifnum [1]{%
 \ifnum #1\expandafter \@firstoftwo
 \else \expandafter \@secondoftwo
 \fi
}%
\providecommand \@ifx [1]{%
 \ifx #1\expandafter \@firstoftwo
 \else \expandafter \@secondoftwo
 \fi
}%
\providecommand \natexlab [1]{#1}%
\providecommand \enquote  [1]{``#1''}%
\providecommand \bibnamefont  [1]{#1}%
\providecommand \bibfnamefont [1]{#1}%
\providecommand \citenamefont [1]{#1}%
\providecommand \href@noop [0]{\@secondoftwo}%
\providecommand \href [0]{\begingroup \@sanitize@url \@href}%
\providecommand \@href[1]{\@@startlink{#1}\@@href}%
\providecommand \@@href[1]{\endgroup#1\@@endlink}%
\providecommand \@sanitize@url [0]{\catcode `\\12\catcode `\$12\catcode
  `\&12\catcode `\#12\catcode `\^12\catcode `\_12\catcode `\%12\relax}%
\providecommand \@@startlink[1]{}%
\providecommand \@@endlink[0]{}%
\providecommand \url  [0]{\begingroup\@sanitize@url \@url }%
\providecommand \@url [1]{\endgroup\@href {#1}{\urlprefix }}%
\providecommand \urlprefix  [0]{URL }%
\providecommand \Eprint [0]{\href }%
\providecommand \doibase [0]{http://dx.doi.org/}%
\providecommand \selectlanguage [0]{\@gobble}%
\providecommand \bibinfo  [0]{\@secondoftwo}%
\providecommand \bibfield  [0]{\@secondoftwo}%
\providecommand \translation [1]{[#1]}%
\providecommand \BibitemOpen [0]{}%
\providecommand \bibitemStop [0]{}%
\providecommand \bibitemNoStop [0]{.\EOS\space}%
\providecommand \EOS [0]{\spacefactor3000\relax}%
\providecommand \BibitemShut  [1]{\csname bibitem#1\endcsname}%
\let\auto@bib@innerbib\@empty
%</preamble>
\bibitem [{\citenamefont {Novoselov}\ \emph {et~al.}(2012)\citenamefont
  {Novoselov}, \citenamefont {Fal'ko}, \citenamefont {Colombo}, \citenamefont
  {Gellert}, \citenamefont {Schwab}, \citenamefont {Kim} \emph
  {et~al.}}]{Novoselov2012}%
  \BibitemOpen
  \bibfield  {author} {\bibinfo {author} {\bibfnamefont {K.}~\bibnamefont
  {Novoselov}}, \bibinfo {author} {\bibfnamefont {V.}~\bibnamefont {Fal'ko}},
  \bibinfo {author} {\bibfnamefont {L.}~\bibnamefont {Colombo}}, \bibinfo
  {author} {\bibfnamefont {P.}~\bibnamefont {Gellert}}, \bibinfo {author}
  {\bibfnamefont {M.}~\bibnamefont {Schwab}}, \bibinfo {author} {\bibfnamefont
  {K.}~\bibnamefont {Kim}},  \emph {et~al.},\ }\href@noop {} {\bibfield
  {journal} {\bibinfo  {journal} {Nature}\ }\textbf {\bibinfo {volume} {490}},\
  \bibinfo {pages} {192} (\bibinfo {year} {2012})}\BibitemShut {NoStop}%
\bibitem [{\citenamefont {Tetlow}\ \emph {et~al.}(2014)\citenamefont {Tetlow},
  \citenamefont {De~Boer}, \citenamefont {Ford}, \citenamefont {Vvedensky},
  \citenamefont {Coraux},\ and\ \citenamefont {Kantorovich}}]{Tetlow2014}%
  \BibitemOpen
  \bibfield  {author} {\bibinfo {author} {\bibfnamefont {H.}~\bibnamefont
  {Tetlow}}, \bibinfo {author} {\bibfnamefont {J.}~\bibnamefont {De~Boer}},
  \bibinfo {author} {\bibfnamefont {I.}~\bibnamefont {Ford}}, \bibinfo {author}
  {\bibfnamefont {D.}~\bibnamefont {Vvedensky}}, \bibinfo {author}
  {\bibfnamefont {J.}~\bibnamefont {Coraux}}, \ and\ \bibinfo {author}
  {\bibfnamefont {L.}~\bibnamefont {Kantorovich}},\ }\href@noop {} {\bibfield
  {journal} {\bibinfo  {journal} {Phys. Rep.}\ }\textbf {\bibinfo {volume}
  {542}},\ \bibinfo {pages} {195} (\bibinfo {year} {2014})}\BibitemShut
  {NoStop}%
\bibitem [{\citenamefont {Pletikosi{\'c}}\ \emph {et~al.}(2009)\citenamefont
  {Pletikosi{\'c}}, \citenamefont {Kralj}, \citenamefont {Pervan},
  \citenamefont {Brako}, \citenamefont {Coraux}, \citenamefont {N'Diaye},
  \citenamefont {Busse},\ and\ \citenamefont {Michely}}]{Pletikosic2009}%
  \BibitemOpen
  \bibfield  {author} {\bibinfo {author} {\bibfnamefont {I.}~\bibnamefont
  {Pletikosi{\'c}}}, \bibinfo {author} {\bibfnamefont {M.}~\bibnamefont
  {Kralj}}, \bibinfo {author} {\bibfnamefont {P.}~\bibnamefont {Pervan}},
  \bibinfo {author} {\bibfnamefont {R.}~\bibnamefont {Brako}}, \bibinfo
  {author} {\bibfnamefont {J.}~\bibnamefont {Coraux}}, \bibinfo {author}
  {\bibfnamefont {A.}~\bibnamefont {N'Diaye}}, \bibinfo {author} {\bibfnamefont
  {C.}~\bibnamefont {Busse}}, \ and\ \bibinfo {author} {\bibfnamefont
  {T.}~\bibnamefont {Michely}},\ }\href@noop {} {\bibfield  {journal} {\bibinfo
   {journal} {Phys. Rev. Lett.}\ }\textbf {\bibinfo {volume} {102}},\ \bibinfo
  {pages} {056808} (\bibinfo {year} {2009})}\BibitemShut {NoStop}%
\bibitem [{\citenamefont {Varykhalov}\ \emph {et~al.}(2008)\citenamefont
  {Varykhalov}, \citenamefont {S{\'a}nchez-Barriga}, \citenamefont {Shikin},
  \citenamefont {Biswas}, \citenamefont {Vescovo}, \citenamefont {Rybkin},
  \citenamefont {Marchenko},\ and\ \citenamefont {Rader}}]{Varykhalov2008}%
  \BibitemOpen
  \bibfield  {author} {\bibinfo {author} {\bibfnamefont {A.}~\bibnamefont
  {Varykhalov}}, \bibinfo {author} {\bibfnamefont {J.}~\bibnamefont
  {S{\'a}nchez-Barriga}}, \bibinfo {author} {\bibfnamefont {A.}~\bibnamefont
  {Shikin}}, \bibinfo {author} {\bibfnamefont {C.}~\bibnamefont {Biswas}},
  \bibinfo {author} {\bibfnamefont {E.}~\bibnamefont {Vescovo}}, \bibinfo
  {author} {\bibfnamefont {A.}~\bibnamefont {Rybkin}}, \bibinfo {author}
  {\bibfnamefont {D.}~\bibnamefont {Marchenko}}, \ and\ \bibinfo {author}
  {\bibfnamefont {O.}~\bibnamefont {Rader}},\ }\href@noop {} {\bibfield
  {journal} {\bibinfo  {journal} {Phys. Rev. Lett.}\ }\textbf {\bibinfo
  {volume} {101}},\ \bibinfo {pages} {157601} (\bibinfo {year}
  {2008})}\BibitemShut {NoStop}%
\bibitem [{\citenamefont {Tonnoir}\ \emph {et~al.}(2013)\citenamefont
  {Tonnoir}, \citenamefont {Kimouche}, \citenamefont {Coraux}, \citenamefont
  {Magaud}, \citenamefont {Delsol}, \citenamefont {Gilles},\ and\ \citenamefont
  {Chapelier}}]{Tonnoir2013}%
  \BibitemOpen
  \bibfield  {author} {\bibinfo {author} {\bibfnamefont {C.}~\bibnamefont
  {Tonnoir}}, \bibinfo {author} {\bibfnamefont {A.}~\bibnamefont {Kimouche}},
  \bibinfo {author} {\bibfnamefont {J.}~\bibnamefont {Coraux}}, \bibinfo
  {author} {\bibfnamefont {L.}~\bibnamefont {Magaud}}, \bibinfo {author}
  {\bibfnamefont {B.}~\bibnamefont {Delsol}}, \bibinfo {author} {\bibfnamefont
  {B.}~\bibnamefont {Gilles}}, \ and\ \bibinfo {author} {\bibfnamefont
  {C.}~\bibnamefont {Chapelier}},\ }\href@noop {} {\bibfield  {journal}
  {\bibinfo  {journal} {Phys. Rev. Lett.}\ }\textbf {\bibinfo {volume} {111}},\
  \bibinfo {pages} {246805} (\bibinfo {year} {2013})}\BibitemShut {NoStop}%
\bibitem [{\citenamefont {Marchini}\ \emph {et~al.}(2007)\citenamefont
  {Marchini}, \citenamefont {G{\"u}nther},\ and\ \citenamefont
  {Wintterlin}}]{Marchini2007}%
  \BibitemOpen
  \bibfield  {author} {\bibinfo {author} {\bibfnamefont {S.}~\bibnamefont
  {Marchini}}, \bibinfo {author} {\bibfnamefont {S.}~\bibnamefont
  {G{\"u}nther}}, \ and\ \bibinfo {author} {\bibfnamefont {J.}~\bibnamefont
  {Wintterlin}},\ }\href@noop {} {\bibfield  {journal} {\bibinfo  {journal}
  {Phys. Rev. B}\ }\textbf {\bibinfo {volume} {76}},\ \bibinfo {pages} {075429}
  (\bibinfo {year} {2007})}\BibitemShut {NoStop}%
\bibitem [{\citenamefont {V{\'a}zquez De~Parga}\ \emph
  {et~al.}(2008)\citenamefont {V{\'a}zquez De~Parga}, \citenamefont {Calleja},
  \citenamefont {Borca}, \citenamefont {Passeggi~Jr}, \citenamefont
  {Hinarejos}, \citenamefont {Guinea},\ and\ \citenamefont
  {Miranda}}]{VazquezdeParga2008}%
  \BibitemOpen
  \bibfield  {author} {\bibinfo {author} {\bibfnamefont {A.}~\bibnamefont
  {V{\'a}zquez De~Parga}}, \bibinfo {author} {\bibfnamefont {F.}~\bibnamefont
  {Calleja}}, \bibinfo {author} {\bibfnamefont {B.}~\bibnamefont {Borca}},
  \bibinfo {author} {\bibfnamefont {M.}~\bibnamefont {Passeggi~Jr}}, \bibinfo
  {author} {\bibfnamefont {J.}~\bibnamefont {Hinarejos}}, \bibinfo {author}
  {\bibfnamefont {F.}~\bibnamefont {Guinea}}, \ and\ \bibinfo {author}
  {\bibfnamefont {R.}~\bibnamefont {Miranda}},\ }\href@noop {} {\bibfield
  {journal} {\bibinfo  {journal} {Phys. Rev. Lett.}\ }\textbf {\bibinfo
  {volume} {100}},\ \bibinfo {pages} {056807} (\bibinfo {year}
  {2008})}\BibitemShut {NoStop}%
\bibitem [{\citenamefont {Busse}\ \emph {et~al.}(2011)\citenamefont {Busse},
  \citenamefont {Lazi{\'c}}, \citenamefont {Djemour}, \citenamefont {Coraux},
  \citenamefont {Gerber}, \citenamefont {Atodiresei}, \citenamefont {Caciuc},
  \citenamefont {Brako}, \citenamefont {Bl{\"u}gel}, \citenamefont {Zegenhagen}
  \emph {et~al.}}]{Busse2011}%
  \BibitemOpen
  \bibfield  {author} {\bibinfo {author} {\bibfnamefont {C.}~\bibnamefont
  {Busse}}, \bibinfo {author} {\bibfnamefont {P.}~\bibnamefont {Lazi{\'c}}},
  \bibinfo {author} {\bibfnamefont {R.}~\bibnamefont {Djemour}}, \bibinfo
  {author} {\bibfnamefont {J.}~\bibnamefont {Coraux}}, \bibinfo {author}
  {\bibfnamefont {T.}~\bibnamefont {Gerber}}, \bibinfo {author} {\bibfnamefont
  {N.}~\bibnamefont {Atodiresei}}, \bibinfo {author} {\bibfnamefont
  {V.}~\bibnamefont {Caciuc}}, \bibinfo {author} {\bibfnamefont
  {R.}~\bibnamefont {Brako}}, \bibinfo {author} {\bibfnamefont
  {S.}~\bibnamefont {Bl{\"u}gel}}, \bibinfo {author} {\bibfnamefont
  {J.}~\bibnamefont {Zegenhagen}},  \emph {et~al.},\ }\href@noop {} {\bibfield
  {journal} {\bibinfo  {journal} {Phys. Rev. Lett.}\ }\textbf {\bibinfo
  {volume} {107}},\ \bibinfo {pages} {036101} (\bibinfo {year}
  {2011})}\BibitemShut {NoStop}%
\bibitem [{\citenamefont {Sun}\ \emph {et~al.}(2011)\citenamefont {Sun},
  \citenamefont {H{\"a}m{\"a}l{\"a}inen}, \citenamefont {Sainio}, \citenamefont
  {Lahtinen}, \citenamefont {Vanmaekelbergh},\ and\ \citenamefont
  {Liljeroth}}]{Sun2011}%
  \BibitemOpen
  \bibfield  {author} {\bibinfo {author} {\bibfnamefont {Z.}~\bibnamefont
  {Sun}}, \bibinfo {author} {\bibfnamefont {S.}~\bibnamefont
  {H{\"a}m{\"a}l{\"a}inen}}, \bibinfo {author} {\bibfnamefont {J.}~\bibnamefont
  {Sainio}}, \bibinfo {author} {\bibfnamefont {J.}~\bibnamefont {Lahtinen}},
  \bibinfo {author} {\bibfnamefont {D.}~\bibnamefont {Vanmaekelbergh}}, \ and\
  \bibinfo {author} {\bibfnamefont {P.}~\bibnamefont {Liljeroth}},\ }\href@noop
  {} {\bibfield  {journal} {\bibinfo  {journal} {Phys. Rev. B}\ }\textbf
  {\bibinfo {volume} {83}},\ \bibinfo {pages} {081415} (\bibinfo {year}
  {2011})}\BibitemShut {NoStop}%
\bibitem [{\citenamefont {Dedkov}\ and\ \citenamefont
  {Voloshina}(2014)}]{Dedkov2014}%
  \BibitemOpen
  \bibfield  {author} {\bibinfo {author} {\bibfnamefont {Y.}~\bibnamefont
  {Dedkov}}\ and\ \bibinfo {author} {\bibfnamefont {E.}~\bibnamefont
  {Voloshina}},\ }\href@noop {} {\bibfield  {journal} {\bibinfo  {journal}
  {Phys. Chem. Chem. Phys.}\ }\textbf {\bibinfo {volume} {16}},\ \bibinfo
  {pages} {3894} (\bibinfo {year} {2014})}\BibitemShut {NoStop}%
\bibitem [{\citenamefont {H{\"a}m{\"a}l{\"a}inen}\ \emph
  {et~al.}(2013)\citenamefont {H{\"a}m{\"a}l{\"a}inen}, \citenamefont
  {Boneschanscher}, \citenamefont {Jacobse}, \citenamefont {Swart},
  \citenamefont {Pussi}, \citenamefont {Moritz}, \citenamefont {Lahtinen},
  \citenamefont {Liljeroth},\ and\ \citenamefont {Sainio}}]{Hamalainen2013}%
  \BibitemOpen
  \bibfield  {author} {\bibinfo {author} {\bibfnamefont {S.}~\bibnamefont
  {H{\"a}m{\"a}l{\"a}inen}}, \bibinfo {author} {\bibfnamefont {M.}~\bibnamefont
  {Boneschanscher}}, \bibinfo {author} {\bibfnamefont {P.}~\bibnamefont
  {Jacobse}}, \bibinfo {author} {\bibfnamefont {I.}~\bibnamefont {Swart}},
  \bibinfo {author} {\bibfnamefont {K.}~\bibnamefont {Pussi}}, \bibinfo
  {author} {\bibfnamefont {W.}~\bibnamefont {Moritz}}, \bibinfo {author}
  {\bibfnamefont {J.}~\bibnamefont {Lahtinen}}, \bibinfo {author}
  {\bibfnamefont {P.}~\bibnamefont {Liljeroth}}, \ and\ \bibinfo {author}
  {\bibfnamefont {J.}~\bibnamefont {Sainio}},\ }\href@noop {} {\bibfield
  {journal} {\bibinfo  {journal} {Phys. Rev. B}\ }\textbf {\bibinfo {volume}
  {88}},\ \bibinfo {pages} {201406} (\bibinfo {year} {2013})}\BibitemShut
  {NoStop}%
\bibitem [{\citenamefont {Moritz}\ \emph {et~al.}(2010)\citenamefont {Moritz},
  \citenamefont {Wang}, \citenamefont {Bocquet}, \citenamefont {Brugger},
  \citenamefont {Greber}, \citenamefont {Wintterlin},\ and\ \citenamefont
  {G{\"u}nther}}]{Moritz2010}%
  \BibitemOpen
  \bibfield  {author} {\bibinfo {author} {\bibfnamefont {W.}~\bibnamefont
  {Moritz}}, \bibinfo {author} {\bibfnamefont {B.}~\bibnamefont {Wang}},
  \bibinfo {author} {\bibfnamefont {M.-L.}\ \bibnamefont {Bocquet}}, \bibinfo
  {author} {\bibfnamefont {T.}~\bibnamefont {Brugger}}, \bibinfo {author}
  {\bibfnamefont {T.}~\bibnamefont {Greber}}, \bibinfo {author} {\bibfnamefont
  {J.}~\bibnamefont {Wintterlin}}, \ and\ \bibinfo {author} {\bibfnamefont
  {S.}~\bibnamefont {G{\"u}nther}},\ }\href@noop {} {\bibfield  {journal}
  {\bibinfo  {journal} {Phys. Rev. Lett.}\ }\textbf {\bibinfo {volume} {104}},\
  \bibinfo {pages} {136102} (\bibinfo {year} {2010})}\BibitemShut {NoStop}%
\bibitem [{\citenamefont {Martoccia}\ \emph {et~al.}(2008)\citenamefont
  {Martoccia}, \citenamefont {Willmott}, \citenamefont {Brugger}, \citenamefont
  {Bj{\"o}rck}, \citenamefont {G{\"u}nther}, \citenamefont {Schlep{\"u}tz},
  \citenamefont {Cervellino}, \citenamefont {Pauli}, \citenamefont {Patterson},
  \citenamefont {Marchini} \emph {et~al.}}]{Martoccia2008}%
  \BibitemOpen
  \bibfield  {author} {\bibinfo {author} {\bibfnamefont {D.}~\bibnamefont
  {Martoccia}}, \bibinfo {author} {\bibfnamefont {P.}~\bibnamefont {Willmott}},
  \bibinfo {author} {\bibfnamefont {T.}~\bibnamefont {Brugger}}, \bibinfo
  {author} {\bibfnamefont {M.}~\bibnamefont {Bj{\"o}rck}}, \bibinfo {author}
  {\bibfnamefont {S.}~\bibnamefont {G{\"u}nther}}, \bibinfo {author}
  {\bibfnamefont {C.}~\bibnamefont {Schlep{\"u}tz}}, \bibinfo {author}
  {\bibfnamefont {A.}~\bibnamefont {Cervellino}}, \bibinfo {author}
  {\bibfnamefont {S.}~\bibnamefont {Pauli}}, \bibinfo {author} {\bibfnamefont
  {B.}~\bibnamefont {Patterson}}, \bibinfo {author} {\bibfnamefont
  {S.}~\bibnamefont {Marchini}},  \emph {et~al.},\ }\href@noop {} {\bibfield
  {journal} {\bibinfo  {journal} {Phys. Rev. Lett.}\ }\textbf {\bibinfo
  {volume} {101}},\ \bibinfo {pages} {126102} (\bibinfo {year}
  {2008})}\BibitemShut {NoStop}%
\bibitem [{\citenamefont {Mittendorfer}\ \emph {et~al.}(2011)\citenamefont
  {Mittendorfer}, \citenamefont {Garhofer}, \citenamefont {Redinger},
  \citenamefont {Klime{\v{s}}}, \citenamefont {Harl},\ and\ \citenamefont
  {Kresse}}]{Mittendorfer2011}%
  \BibitemOpen
  \bibfield  {author} {\bibinfo {author} {\bibfnamefont {F.}~\bibnamefont
  {Mittendorfer}}, \bibinfo {author} {\bibfnamefont {A.}~\bibnamefont
  {Garhofer}}, \bibinfo {author} {\bibfnamefont {J.}~\bibnamefont {Redinger}},
  \bibinfo {author} {\bibfnamefont {J.}~\bibnamefont {Klime{\v{s}}}}, \bibinfo
  {author} {\bibfnamefont {J.}~\bibnamefont {Harl}}, \ and\ \bibinfo {author}
  {\bibfnamefont {G.}~\bibnamefont {Kresse}},\ }\href@noop {} {\bibfield
  {journal} {\bibinfo  {journal} {Phys. Rev. B}\ }\textbf {\bibinfo {volume}
  {84}},\ \bibinfo {pages} {201401} (\bibinfo {year} {2011})}\BibitemShut
  {NoStop}%
\bibitem [{\citenamefont {Jean}\ \emph {et~al.}(2013)\citenamefont {Jean},
  \citenamefont {Zhou}, \citenamefont {Blanc}, \citenamefont {Felici},
  \citenamefont {Coraux},\ and\ \citenamefont {Renaud}}]{Jean2013}%
  \BibitemOpen
  \bibfield  {author} {\bibinfo {author} {\bibfnamefont {F.}~\bibnamefont
  {Jean}}, \bibinfo {author} {\bibfnamefont {T.}~\bibnamefont {Zhou}}, \bibinfo
  {author} {\bibfnamefont {N.}~\bibnamefont {Blanc}}, \bibinfo {author}
  {\bibfnamefont {R.}~\bibnamefont {Felici}}, \bibinfo {author} {\bibfnamefont
  {J.}~\bibnamefont {Coraux}}, \ and\ \bibinfo {author} {\bibfnamefont
  {G.}~\bibnamefont {Renaud}},\ }\href@noop {} {\bibfield  {journal} {\bibinfo
  {journal} {Phys. Rev. B}\ }\textbf {\bibinfo {volume} {88}},\ \bibinfo
  {pages} {165406} (\bibinfo {year} {2013})}\BibitemShut {NoStop}%
\bibitem [{\citenamefont {Drnec}\ \emph {et~al.}(2014)\citenamefont {Drnec},
  \citenamefont {Zhou}, \citenamefont {Pintea}, \citenamefont {Onderwaater},
  \citenamefont {Vlieg}, \citenamefont {Renaud},\ and\ \citenamefont
  {Felici}}]{Drnec2014}%
  \BibitemOpen
  \bibfield  {author} {\bibinfo {author} {\bibfnamefont {J.}~\bibnamefont
  {Drnec}}, \bibinfo {author} {\bibfnamefont {T.}~\bibnamefont {Zhou}},
  \bibinfo {author} {\bibfnamefont {S.}~\bibnamefont {Pintea}}, \bibinfo
  {author} {\bibfnamefont {W.}~\bibnamefont {Onderwaater}}, \bibinfo {author}
  {\bibfnamefont {E.}~\bibnamefont {Vlieg}}, \bibinfo {author} {\bibfnamefont
  {G.}~\bibnamefont {Renaud}}, \ and\ \bibinfo {author} {\bibfnamefont
  {R.}~\bibnamefont {Felici}},\ }\href@noop {} {\bibfield  {journal} {\bibinfo
  {journal} {J. Appl. Crystallogr.}\ }\textbf {\bibinfo {volume} {47}},\
  \bibinfo {pages} {365} (\bibinfo {year} {2014})}\BibitemShut {NoStop}%
\bibitem [{\citenamefont {Van~Gastel}\ \emph {et~al.}(2009)\citenamefont
  {Van~Gastel}, \citenamefont {N’Diaye}, \citenamefont {Wall}, \citenamefont
  {Coraux}, \citenamefont {Busse}, \citenamefont {Buckanie}, \citenamefont
  {Meyer Zu~Heringdorf}, \citenamefont {Horn Von~Hoegen}, \citenamefont
  {Michely},\ and\ \citenamefont {Poelsema}}]{vanGastel2009}%
  \BibitemOpen
  \bibfield  {author} {\bibinfo {author} {\bibfnamefont {R.}~\bibnamefont
  {Van~Gastel}}, \bibinfo {author} {\bibfnamefont {A.}~\bibnamefont
  {N’Diaye}}, \bibinfo {author} {\bibfnamefont {D.}~\bibnamefont {Wall}},
  \bibinfo {author} {\bibfnamefont {J.}~\bibnamefont {Coraux}}, \bibinfo
  {author} {\bibfnamefont {C.}~\bibnamefont {Busse}}, \bibinfo {author}
  {\bibfnamefont {N.}~\bibnamefont {Buckanie}}, \bibinfo {author}
  {\bibfnamefont {F.-J.}\ \bibnamefont {Meyer Zu~Heringdorf}}, \bibinfo
  {author} {\bibfnamefont {M.}~\bibnamefont {Horn Von~Hoegen}}, \bibinfo
  {author} {\bibfnamefont {T.}~\bibnamefont {Michely}}, \ and\ \bibinfo
  {author} {\bibfnamefont {B.}~\bibnamefont {Poelsema}},\ }\href@noop {}
  {\bibfield  {journal} {\bibinfo  {journal} {Applied Physics Letters}\
  }\textbf {\bibinfo {volume} {95}},\ \bibinfo {pages} {121901} (\bibinfo
  {year} {2009})}\BibitemShut {NoStop}%
\bibitem [{\citenamefont {Runte}\ \emph {et~al.}(2014)\citenamefont {Runte},
  \citenamefont {Lazi{\'c}}, \citenamefont {Vo-Van}, \citenamefont {Coraux},
  \citenamefont {Zegenhagen},\ and\ \citenamefont {Busse}}]{Runte2014}%
  \BibitemOpen
  \bibfield  {author} {\bibinfo {author} {\bibfnamefont {S.}~\bibnamefont
  {Runte}}, \bibinfo {author} {\bibfnamefont {P.}~\bibnamefont {Lazi{\'c}}},
  \bibinfo {author} {\bibfnamefont {C.}~\bibnamefont {Vo-Van}}, \bibinfo
  {author} {\bibfnamefont {J.}~\bibnamefont {Coraux}}, \bibinfo {author}
  {\bibfnamefont {J.}~\bibnamefont {Zegenhagen}}, \ and\ \bibinfo {author}
  {\bibfnamefont {C.}~\bibnamefont {Busse}},\ }\href@noop {} {\bibfield
  {journal} {\bibinfo  {journal} {Phys. Rev. B}\ }\textbf {\bibinfo {volume}
  {89}},\ \bibinfo {pages} {155427} (\bibinfo {year} {2014})}\BibitemShut
  {NoStop}%
\bibitem [{\citenamefont {Blanc}\ \emph {et~al.}(2012)\citenamefont {Blanc},
  \citenamefont {Coraux}, \citenamefont {Vo-Van}, \citenamefont {Geaymond},
  \citenamefont {Renaud} \emph {et~al.}}]{Blanc2012}%
  \BibitemOpen
  \bibfield  {author} {\bibinfo {author} {\bibfnamefont {N.}~\bibnamefont
  {Blanc}}, \bibinfo {author} {\bibfnamefont {J.}~\bibnamefont {Coraux}},
  \bibinfo {author} {\bibfnamefont {C.}~\bibnamefont {Vo-Van}}, \bibinfo
  {author} {\bibfnamefont {O.}~\bibnamefont {Geaymond}}, \bibinfo {author}
  {\bibfnamefont {G.}~\bibnamefont {Renaud}},  \emph {et~al.},\ }\href@noop {}
  {\bibfield  {journal} {\bibinfo  {journal} {Phys. Rev. B}\ }\textbf {\bibinfo
  {volume} {86}},\ \bibinfo {pages} {235439} (\bibinfo {year}
  {2012})}\BibitemShut {NoStop}%
\bibitem [{\citenamefont {Usachov}\ \emph {et~al.}(2012)\citenamefont
  {Usachov}, \citenamefont {Fedorov}, \citenamefont {Vilkov}, \citenamefont
  {Adamchuk}, \citenamefont {Yashina}, \citenamefont {Bondarenko},
  \citenamefont {Saranin}, \citenamefont {Gr{\"u}neis},\ and\ \citenamefont
  {Vyalikh}}]{Usachov2012}%
  \BibitemOpen
  \bibfield  {author} {\bibinfo {author} {\bibfnamefont {D.}~\bibnamefont
  {Usachov}}, \bibinfo {author} {\bibfnamefont {A.}~\bibnamefont {Fedorov}},
  \bibinfo {author} {\bibfnamefont {O.}~\bibnamefont {Vilkov}}, \bibinfo
  {author} {\bibfnamefont {V.}~\bibnamefont {Adamchuk}}, \bibinfo {author}
  {\bibfnamefont {L.}~\bibnamefont {Yashina}}, \bibinfo {author} {\bibfnamefont
  {L.}~\bibnamefont {Bondarenko}}, \bibinfo {author} {\bibfnamefont
  {A.}~\bibnamefont {Saranin}}, \bibinfo {author} {\bibfnamefont
  {A.}~\bibnamefont {Gr{\"u}neis}}, \ and\ \bibinfo {author} {\bibfnamefont
  {D.}~\bibnamefont {Vyalikh}},\ }\href@noop {} {\bibfield  {journal} {\bibinfo
   {journal} {Phys. Rev. B}\ }\textbf {\bibinfo {volume} {86}},\ \bibinfo
  {pages} {155151} (\bibinfo {year} {2012})}\BibitemShut {NoStop}%
\bibitem [{\citenamefont {Coraux}\ \emph {et~al.}(2008)\citenamefont {Coraux},
  \citenamefont {N'Diaye}, \citenamefont {Busse},\ and\ \citenamefont
  {Michely}}]{Coraux2008}%
  \BibitemOpen
  \bibfield  {author} {\bibinfo {author} {\bibfnamefont {J.}~\bibnamefont
  {Coraux}}, \bibinfo {author} {\bibfnamefont {A.}~\bibnamefont {N'Diaye}},
  \bibinfo {author} {\bibfnamefont {C.}~\bibnamefont {Busse}}, \ and\ \bibinfo
  {author} {\bibfnamefont {T.}~\bibnamefont {Michely}},\ }\href@noop {}
  {\bibfield  {journal} {\bibinfo  {journal} {Nano Lett.}\ }\textbf {\bibinfo
  {volume} {8}},\ \bibinfo {pages} {565} (\bibinfo {year} {2008})}\BibitemShut
  {NoStop}%
\bibitem [{\citenamefont {Charrier}\ \emph {et~al.}(2002)\citenamefont
  {Charrier}, \citenamefont {Coati}, \citenamefont {Argunova}, \citenamefont
  {Thibaudau}, \citenamefont {Garreau}, \citenamefont {Pinchaux}, \citenamefont
  {Forbeaux}, \citenamefont {Debever}, \citenamefont {Sauvage-Simkin},\ and\
  \citenamefont {Themlin}}]{Charrier2002}%
  \BibitemOpen
  \bibfield  {author} {\bibinfo {author} {\bibfnamefont {A.}~\bibnamefont
  {Charrier}}, \bibinfo {author} {\bibfnamefont {A.}~\bibnamefont {Coati}},
  \bibinfo {author} {\bibfnamefont {T.}~\bibnamefont {Argunova}}, \bibinfo
  {author} {\bibfnamefont {F.}~\bibnamefont {Thibaudau}}, \bibinfo {author}
  {\bibfnamefont {Y.}~\bibnamefont {Garreau}}, \bibinfo {author} {\bibfnamefont
  {R.}~\bibnamefont {Pinchaux}}, \bibinfo {author} {\bibfnamefont
  {I.}~\bibnamefont {Forbeaux}}, \bibinfo {author} {\bibfnamefont {J.-M.}\
  \bibnamefont {Debever}}, \bibinfo {author} {\bibfnamefont {M.}~\bibnamefont
  {Sauvage-Simkin}}, \ and\ \bibinfo {author} {\bibfnamefont {J.-M.}\
  \bibnamefont {Themlin}},\ }\href@noop {} {\bibfield  {journal} {\bibinfo
  {journal} {J. Appl. Phys.}\ }\textbf {\bibinfo {volume} {92}},\ \bibinfo
  {pages} {2479} (\bibinfo {year} {2002})}\BibitemShut {NoStop}%
\bibitem [{Not({\natexlab{a}})}]{Note1}%
  \BibitemOpen
  \href@noop {} {} ({\natexlab{a}}),\ \bibinfo {note} {independently refining
  the positions of each atom in a commensurate cell, comprising few
  100$^\circ$C and Ir atoms, obviously would provide a non reliable structural
  picture given that the number of free parameters would approach or even
  exceed the number of experimental points.}\BibitemShut {Stop}%
\bibitem [{\citenamefont {Martoccia}\ \emph {et~al.}(2010)\citenamefont
  {Martoccia}, \citenamefont {Bj{\"o}rck}, \citenamefont {Schlep{\"u}tz},
  \citenamefont {Brugger}, \citenamefont {Pauli}, \citenamefont {Patterson},
  \citenamefont {Greber},\ and\ \citenamefont {Willmott}}]{Martoccia2010}%
  \BibitemOpen
  \bibfield  {author} {\bibinfo {author} {\bibfnamefont {D.}~\bibnamefont
  {Martoccia}}, \bibinfo {author} {\bibfnamefont {M.}~\bibnamefont
  {Bj{\"o}rck}}, \bibinfo {author} {\bibfnamefont {C.}~\bibnamefont
  {Schlep{\"u}tz}}, \bibinfo {author} {\bibfnamefont {T.}~\bibnamefont
  {Brugger}}, \bibinfo {author} {\bibfnamefont {S.}~\bibnamefont {Pauli}},
  \bibinfo {author} {\bibfnamefont {B.}~\bibnamefont {Patterson}}, \bibinfo
  {author} {\bibfnamefont {T.}~\bibnamefont {Greber}}, \ and\ \bibinfo {author}
  {\bibfnamefont {P.}~\bibnamefont {Willmott}},\ }\href@noop {} {\bibfield
  {journal} {\bibinfo  {journal} {New J. Phys.}\ }\textbf {\bibinfo {volume}
  {12}},\ \bibinfo {pages} {043028} (\bibinfo {year} {2010})}\BibitemShut
  {NoStop}%
\bibitem [{\citenamefont {Adams}\ \emph {et~al.}(1985)\citenamefont {Adams},
  \citenamefont {Petersen},\ and\ \citenamefont {Sorensen}}]{Adams1985}%
  \BibitemOpen
  \bibfield  {author} {\bibinfo {author} {\bibfnamefont {D.}~\bibnamefont
  {Adams}}, \bibinfo {author} {\bibfnamefont {L.}~\bibnamefont {Petersen}}, \
  and\ \bibinfo {author} {\bibfnamefont {C.}~\bibnamefont {Sorensen}},\
  }\href@noop {} {\bibfield  {journal} {\bibinfo  {journal} {J. Phys. C}\
  }\textbf {\bibinfo {volume} {18}},\ \bibinfo {pages} {1753} (\bibinfo {year}
  {1985})}\BibitemShut {NoStop}%
\bibitem [{Not({\natexlab{b}})}]{Note2}%
  \BibitemOpen
  \href@noop {} {} ({\natexlab{b}}),\ \bibinfo {note} {actually, we tested that
  even 0.05 \AA \space in-plane displacements have no substantial
  effect.}\BibitemShut {Stop}%
\bibitem [{\citenamefont {Robinson}(1986)}]{Robinson1986}%
  \BibitemOpen
  \bibfield  {author} {\bibinfo {author} {\bibfnamefont {I.}~\bibnamefont
  {Robinson}},\ }\href@noop {} {\bibfield  {journal} {\bibinfo  {journal}
  {Phys. Rev. B}\ }\textbf {\bibinfo {volume} {33}},\ \bibinfo {pages} {3830}
  (\bibinfo {year} {1986})}\BibitemShut {NoStop}%
\bibitem [{\citenamefont {Boneschanscher}\ \emph {et~al.}(2012)\citenamefont
  {Boneschanscher}, \citenamefont {van~der Lit}, \citenamefont {Sun},
  \citenamefont {Swart}, \citenamefont {Liljeroth},\ and\ \citenamefont
  {Vanmaekelbergh}}]{Boneschanscher2012}%
  \BibitemOpen
  \bibfield  {author} {\bibinfo {author} {\bibfnamefont {M.}~\bibnamefont
  {Boneschanscher}}, \bibinfo {author} {\bibfnamefont {J.}~\bibnamefont
  {van~der Lit}}, \bibinfo {author} {\bibfnamefont {Z.}~\bibnamefont {Sun}},
  \bibinfo {author} {\bibfnamefont {I.}~\bibnamefont {Swart}}, \bibinfo
  {author} {\bibfnamefont {P.}~\bibnamefont {Liljeroth}}, \ and\ \bibinfo
  {author} {\bibfnamefont {D.}~\bibnamefont {Vanmaekelbergh}},\ }\href@noop {}
  {\bibfield  {journal} {\bibinfo  {journal} {ACS Nano}\ }\textbf {\bibinfo
  {volume} {6}},\ \bibinfo {pages} {10216} (\bibinfo {year}
  {2012})}\BibitemShut {NoStop}%
\bibitem [{\citenamefont {N'Diaye}\ \emph {et~al.}(2006)\citenamefont
  {N'Diaye}, \citenamefont {Bleikamp}, \citenamefont {Feibelman},\ and\
  \citenamefont {Michely}}]{Ndiaye2006}%
  \BibitemOpen
  \bibfield  {author} {\bibinfo {author} {\bibfnamefont {A.}~\bibnamefont
  {N'Diaye}}, \bibinfo {author} {\bibfnamefont {S.}~\bibnamefont {Bleikamp}},
  \bibinfo {author} {\bibfnamefont {P.}~\bibnamefont {Feibelman}}, \ and\
  \bibinfo {author} {\bibfnamefont {T.}~\bibnamefont {Michely}},\ }\href@noop
  {} {\bibfield  {journal} {\bibinfo  {journal} {Phys. Rev. Lett.}\ }\textbf
  {\bibinfo {volume} {97}},\ \bibinfo {pages} {215501} (\bibinfo {year}
  {2006})}\BibitemShut {NoStop}%
\bibitem [{\citenamefont {He}\ \emph {et~al.}(2008)\citenamefont {He},
  \citenamefont {Stierle}, \citenamefont {Li}, \citenamefont {Farkas},
  \citenamefont {Kasper},\ and\ \citenamefont {Over}}]{He2008}%
  \BibitemOpen
  \bibfield  {author} {\bibinfo {author} {\bibfnamefont {Y.}~\bibnamefont
  {He}}, \bibinfo {author} {\bibfnamefont {A.}~\bibnamefont {Stierle}},
  \bibinfo {author} {\bibfnamefont {W.}~\bibnamefont {Li}}, \bibinfo {author}
  {\bibfnamefont {A.}~\bibnamefont {Farkas}}, \bibinfo {author} {\bibfnamefont
  {N.}~\bibnamefont {Kasper}}, \ and\ \bibinfo {author} {\bibfnamefont
  {H.}~\bibnamefont {Over}},\ }\href@noop {} {\bibfield  {journal} {\bibinfo
  {journal} {J. Phys. Chem. C}\ }\textbf {\bibinfo {volume} {112}},\ \bibinfo
  {pages} {11946} (\bibinfo {year} {2008})}\BibitemShut {NoStop}%
\end{thebibliography}
\end{document}